# Deep Generative Models-Assisted Automated Labeling for Electron Microscopy Images Segmentation

Wenhao Yuan[1], Bingqing Yao[1], Shengdong Tan[1], Fengqi You[2*], Qian He[1, 3*]

[1] Department of Material Science and Engineering, College of Design and Engineering, National University of Singapore, 9 Engineering Drive 1, EA #03-09, 117575, Singapore

[2] Systems Engineering, Cornell University, Ithaca, NY 14853, USA

[3] Centre for Hydrogen Innovations, National University of Singapore, E8, 1 Engineering Drive 3, 117580, Singapore

* Corresponding author: heqian@nus.edu.sg, fengqi.you@cornell.edu

**Abstract:** The rapid advancement of deep learning has facilitated the automated processing of electron microscopy (EM) big data stacks. However, designing a framework that eliminates manual labeling and adapts to domain gaps remains challenging. Current research remains entangled in the dilemma of pursuing complete automation while still requiring simulations or slight manual annotations. Here we demonstrate tandem generative adversarial network (tGAN), a fully label-free and simulation-free pipeline capable of generating EM images for computer vision training. The tGAN can assimilate key features from new data stacks, thus producing a tailored virtual dataset for the training of automated EM analysis tools. Using segmenting nanoparticles for analyzing size distribution of supported catalysts as the demonstration, our findings showcased that the recognition accuracy of tGAN even exceeds the manually-labeling method by 5%. It can also be adaptively deployed to various data domains without further manual manipulation, which is verified by transfer learning from HAADF-STEM to BF-TEM. This generalizability may enable it to extend its application to a broader range of imaging characterizations, liberating microscopists and materials scientists from tedious dataset annotations.

# Introduction

Electron microscopy (EM) plays a pivotal role in various modern technological sectors[1, 2], underpinning advancements in materials relevant to quantum computing[3], energy[4], and healthcare[5]. Conventionally, the analysis of EM images is conducted manually by researchers utilizing image analysis software such as ImageJ[6] and cisTEM[7]. However, this manual approach is prone to human errors, subjective inconsistency, time inefficiency, and limited scalability concerning large volumes of data.[8] Due to the exponential growth in data-production rates[9-11], scalability has become a critical long-term issue. To address these challenges, there is an urgent need for automated tools capable of efficiently analyzing this burgeoning big data, thereby accelerating the distill of vast multidimensional datasets into meaningful descriptors linked to underlying physical models[9].

The rapid advancement of deep learning (DL) approaches originally developed for computer vision has facilitated the automated analysis of EM images.[8, 12, 13] The two most representative, objective detection and semantic segmentation have been widely used in various applications such as atom/defect detection[14-18], nanoparticle identification[19-21], and crystal structure classification[22-24], etc. However, a major challenge in building such supervised models is that requires sufficient experimental data paired with ground truth for training[13], which is both time- and resource-intensive. Although few-shot learning has been implemented for scanning transmission electron microscope (STEM) image segmentation[24], it still requires manual labeling, limiting its practical application due to the need for expert knowledge and the time-consuming nature of the process. Another common challenge for pre-trained DL models is the failure when facing domain gap[25] (*i.e.*, distribution shift[2]) due to its well-documented brittleness[26]. That is, generalizing the model trained under a set of parameters (*e.g.*, acquisition parameters, sample conditions) to parameters outside the training range is a challenge.[2] Together, these bottlenecks have led to compromising the generalizability of DL models for EM image analysis[2], hindering the large-scale popularization of automated tools in this field.

One promising alternative to labeled experimental data is in silico-generated data, *i.e.*, synthetic training data from physics-based simulation and deep learning-based generative models.[8] Typically, for most cases with simple image patterns (*e.g.*, atomic-scale images), physics-based simulation is an effective method for generating labeled images.[14, 15, 18, 27] For those complex scenarios with poor simulation results (*e.g.*, nano/micro-scale images), deep generative models have been proved to be a good implementation of simulation-to-reality (Sim2Real) domain transfer, which can serve as a relay to further make the simulated images more realistic. Ma *et al.*[28] employed a transfer learning strategy, using a conditional GAN (pix2pix) to make the simulated optical microscopy (OM) images more realistic for the training of polycrystalline iron segmentation. Khan *et al.*[29] utilized a cycle generative adversarial network (CycleGAN) with a reciprocal space discriminator to minimize the difference between simulated and experimental STEM data. Bals *et al.*[30] demonstrated that particle assemblies created by Blender can be converted into artificial scanning electron microscopy (SEM) images with a CycleGAN. Zhang *et al.*[31] also utilized CycleGAN to implement Sim2Real transfer for microrobot on both SEM and OM images. In addition to Sim2Real transfer, some works had shown that image data can also be generated directly from deep generative models (*e.g.*, using StyleGAN2-ada to generate single-cell OM images[32] and StyleGAN3 to generate 2D materials OM images[33]), but they still have not achieved complete label-free due to their basis of supervision concept. These examples showcased the potential of generating synthetic data. However, for simulation-based methods, the variance of multiple physical and chemical factors can dramatically affect the results generated by simulations, and the simulations themselves remain time-consuming and inefficient[13], both of which still inevitably limit the fast generalizability of DL models in EM applications.

In the field of heterogeneous catalysis, the particle size distribution (PSD) of supported nanoparticles are key parameters for interpreting catalytic performance and sintering mechanisms and are of critical importance to both microscopists and materials scientists.[34-37] However, STEM images of these catalysts generally have different morphology and contrast. Thus, developing tools with fast generalizability to analyze various supported nanoparticles will provide useful guidance for developing sinter-resistant catalysts in the industrial production of chemicals.[38, 39]

In this work, we constructed a tandem generative adversarial network (tGAN) pipeline to generate reasonable EM images while simultaneously achieving both label-free and simulation-free, which, to the best of our knowledge, has not been done before. We will showcase that due to the tandem pipeline design that can successively extract morphology and contrast features, tGAN can provide superior segmentation inference compared to manual-labeling method. As an example of the application, the synthetic data was utilized for training a nanoparticles segmentation network (NPsSegNet) on both high-angle annular dark-field (HAADF) STEM images and bright field transmission electron microscope (BF-TEM) images, segmenting nanoparticles and providing PSD information of catalysts. Taking a step further, we have developed it into a co-pilot, named EMcopilot. EMcopilot connects a powerful GPU to the microscope computer via high-speed data communication and uses automated scripts to achieve real-time data capturing and sharing, which can perform computer vision analysis for EM within a response delay of seconds, providing on-the-fly analysis and real-time feedback for *in-situ* experiments.

## Methods

**Experimental STEM Imaging**

PdSn@Al$_2$O$_3$ samples were imaged in a JEOL ARM-200CF instrument operated at 200 kV on the scanning mode, at magnifications between 2,500,000–5,000,000X. Images were acquired at a resolution of 2,048 × 2,048 pixels with a dimension range of 38.438 nm$^2$ –76.876 nm$^2$, resulting in a final sampling size of 0.038–0.019 nm·pixel$^{-1}$.

**Data Preprocessing**

To reduce computational demands while maintaining image detail, we resized each image to 512 × 512 pixels. For training dataset, these resized images were augmented using Flip, ShiftScaleRotate, and GaussNoise from Albumentations[40] successively, all with a probability value of 0.5. For testing dataset, it was only normalized using Albumentations without data augmentation. The ratio of training and testing dataset was set to 4:1. More detailed information regarding subset size, diversity and preprocessing steps for each models has been provided in **Table S1**.

**Nanoparticles Segmentation Network (NPsSegNet) training**

As the Unet based model has been recently widely validated to be the most effective for segmenting EM images[8, 30, 41], the NPsSegNet was implemented based on the UNet++ architecture developed by Zhou et al.[42], following the same encoder-decoder structure. We modified the standard UNet++ structure for accurate segmentation and to prevent overfitting. Training was performed on a NVIDIA GeForce RTX 2070 SUPER GPU under PyTorch CUDA acceleration. The encoder used for UNet++ is ResNet34[43].

The model used FocalLoss[44] (the mode uses multiclass with the alpha of 0.25) and JaccardLoss[45] (the mode uses multiclass) for the loss functions and AdamW[46] as the optimizer with a learning rate of $3 \times 10^{-4}$ and a weight decay of $5 \times 10^{-4}$. StepLR was used the learning scheduler with a step size of 5, gamma of 0.9. The model was limited to training for a maximum of 150 epochs and the batch size was set to 4.

**Segmentation Evaluation**

NPsSegNet classifies EM image pixels into nanoparticles and background, achieving binary semantic segmentation. We evaluated performance using metrics such as Pixel Accuracy (PA) and Mean Intersection over Union (MIoU), which are critical for assessing model precision and overall segmentation quality. Therefore, we defined the true positive (TP) as the number of pixels where NPsSegNet correctly predicts positive examples. The true negative (TN) is defined as the number of pixels where NPsSegNet correctly predicts negative examples. The false positive (FP) is defined as the number of pixels where NPsSegNet predicts positive examples as negative examples. The false negative (FN) is defined as the number of pixels where NPsSegNet predicts negative examples as positive examples. To evaluate the segmentation performance, we adopted there popular metrics: (1) Loss, (2) Pixel Accuracy (PA), and (3) Mean Intersection over Union (MIoU). Loss is the difference between prediction and ground truth, determined by averaged values of FocalLoss and JaccardLoss. PA is the ratio of the number of pixels with the correct predicted class to the total number of pixels, which is calculated as the sum of diagonal elements in the confusion matrix divided by the sum of all elements. MIoU is the ratio of the intersection and union of the model's predicted results for each class and the true value, summed and averaged.

**GANs training**

Both CycleGAN and Pix2Pix models were trained for 100 epochs with an additional 50 epochs for learning rate decay, using a batch size of 4 for training and 1 for testing. Hyperparameter tuning determined optimal values for learning rate, batch size, and the L1 term weight (lambda set to 10). The initial learning rate was set to 0.0002, with a lambda learning rate policy and a decay every 50 epochs. The Adam optimizer was used with beta1 set to 0.5 and beta2 set to 0.999. The generator and the discriminator were both based on a U-Net++ architecture. Loss functions included GAN loss, L1 loss, and MSE loss for stability. The training loop updated the discriminator and generator alternately. Convergence was monitored through loss stabilization and visual fidelity of generated images. Training was also performed on a NVIDIA GeForce RTX 2070 SUPER GPU under PyTorch CUDA acceleration. It is worth noting that unlike NPsSegNet, the Loss of GANs is constantly fluctuating during the training process, and the convergence depends on the quality and fidelity of the generated images. So, the best model is generally considered to be the one with the best resulting image.

**GANs evaluation**

Four popular metrics for GANs, Fréchet inception distance (FID), inception score (IS), structural similarity index measure (SSIM), and peak signal to noise ratio (PSNR), were adopted to evaluate the quality of image data generated by tGAN. FID extracts the image features through the pre-trained inception network, calculates the mean and covariance matrices of the feature distributions of the generated and real images, and quantifies the difference in distribution between the two using the Fréchet distance. The lower the FID, the more realistic the generated image is. IS based on the categorical probability distribution of the generated images, the clarity and diversity of the image is evaluated by calculating the information entropy of the categorical distribution of each image and the Kullback–Leibler divergence of the average categorical distribution of all the images. The higher the IS value, the better the diversity of the generated image. SSIM compares the luminance, contrast, and structure between the generated and reference images, with higher SSIM values indicating greater structural similarity. PSNR of an image is assessed by calculating the mean square error between the generated and real images, with higher values indicating better image quality. Maximum mean discrepancy (MMD) was also used to evaluate the quality of generated images. MMD compares the distributions of real and generated images by embedding them into a reproducing kernel Hilbert space and computing the difference between their distributions using a kernel function. Lower MMD values indicate that the generated images are closer to the real images in terms of distribution.

## Results and Discussion

**Data Annotation Concerns**
**Figure 1** summarizes how our tGAN achieves both simulation-free and label-free. As shown in **Figure 1a**, for the field of heterogeneous catalysis, variations in both the catalyst itself and the characterisation conditions can greatly affect the patterns of EM images. The resulting data stacks are often dispersed over diverse domains with various materials morphology and imaging contrast (**Figure 1b**), placing high demands on the generalizability of the automated analysis tools. However, one single round of annotation using the conventional method of manual labeling can only train one model applicable under a specific domain. Meanwhile, due to the inaccuracy of the ground truth caused by human bias and the inefficiency when facing big data caused by the time-consuming nature of manual-labeling, annotating all the data is infeasible. Thus, utilizing conventional method to deal with the diverse data stack will inevitably waste a large amount of data (*i.e.*, unlabeled large subset in the Figure), which in turn leads to underutilization of the data.

**Tandem GANs Design**
With these two concerns in mind, we designed tGAN with a tandem architecture to achieve full utilization of the dataset without manual-labeling. We deconstructed typical STEM images into morphology and contrast features (**Figure 1b**). The first GAN model (pix2pix) was trained on a subset to capture morphological features, while the second GAN model (CycleGAN) was trained to refine contrast features, ensuring accurate representation of real-world data. The experimentally unlabeled data is then divided into two subsets used to capture these two features separately. Considering that the morphology feature is more accessible compared to the contrast, the smaller subset is used to decipher the morphology information (**Figure 1c**), and subsequently, the larger subset is used to decipher the contrast information (**Figure 1d**). Both processes of deciphering the information are a process of training the GAN to learn the image pattern. Eventually two trained GANs connected in series will fully grasp the image features of the current domain and can generate virtual data for training NPsSegNet (**Figure 1f**).

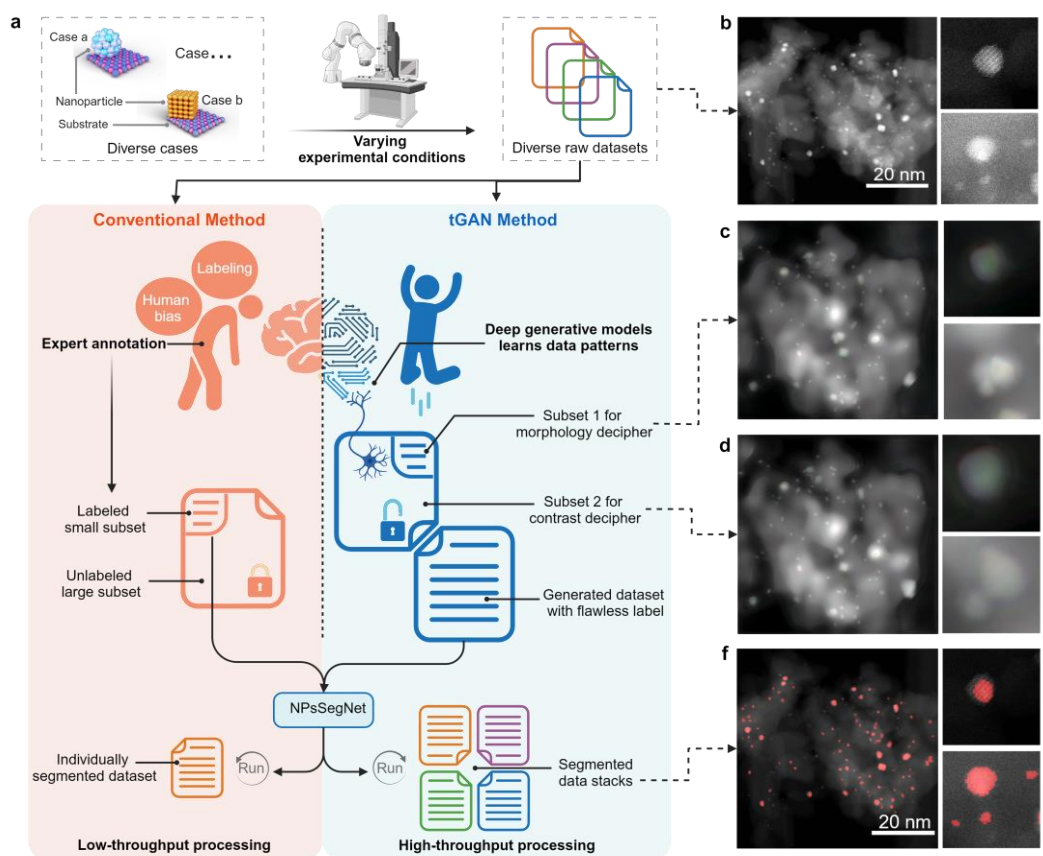

**Figure 1 | Differences between conventional data labeling methods and the tGAN method, with the steps involved in the tandem architecture highlighted.** (a) Comparison of data labeling workflow between conventional method and tGAN. (b) Representative HAADF image of raw data, with magnified images of the two selected areas on the right. (c) Representative generated image after morphology decipher. (c) Representative generated image after contrast decipher. (c) Representative segmentation result under tGAN approach.

In order to elaborate the data generation process of tGAN more clearly, its workflow is shown in **Figure 2a**. As mentioned above, the original unlabeled HAADF-STEM dataset (**Figure 2b**) will be divided into two subsets (step 1), where subset 1 will be sent to the pre-trained model as input for rough segmentation to obtain the rough results (step 2). Since the purpose of this step is to obtain rough morphological information, such as the size, shape, and distribution pattern, etc., the high-precision segmentation results are not strictly needed, *i.e.*, the presence of distribution shift will not affect the subsequent results, which will be verified later. In general, the networks that have been reported in the field, including Segment Anything Model (SAM)[47], are capable of achieving this goal. The obtained rough results are then used as ground truth for the training of the first conditional GAN (pix2pix[48]) together with subset 1 (step 3). The pix2pix model is chosen here because paired image translation are necessary to capture the morphological information of the images. At this point, the trained pix2pix is capable of labels-to-images translation. Therefore, by feeding randomly generated masks (**Figure 2c**) to pix2pix (step 4), we can obtain a large amount of virtual data, i.e., the 1st generation (**Figure 2d**), which just has the topographical characteristics of the real data.

The next step is to decipher the contrast features. The 1st generation obtain from pix2pix and the remaining subset 2 are used to train the second GAN (CycleGAN[49]), which gives CycleGAN the ability to perform a Sim2Real-like translation (step 5), allowing it to learn the contrast information in the images. At this point, the 1st generation from the previous step is sent to CycleGAN as input (step 6), and the output 2nd generation (**Figure 2e**) will become more realistic because of the richer contrast information. At this point, the tGAN formed by pix2pix and CycleGAN in tandem has fully utilized the entire dataset to learn the morphology and contrast features and generated realistic STEM images without the need of manual labeling or simulation. Finally, these generated virtual data will be used to train the NPsSegNet (step 7) and used in the semantic segmentation task (**Figure 2f**) to replace the original pre-trained model at step 2 and thus achieve model adaptation.

**Handling of Domain Gap:**

To further validate the extensibility and generalization of the tGAN pipeline, we purposely observe the performance of tGAN in the face of domain gap transformed from HAADF-STEM to BF-TEM. The pre-trained NPsSegNet using HAADF-STEM images was applied to step 2, rough segmentation of the BF-TEM images process. The intermediate and final results of running the whole pipeline are shown in **Figure 2g-k**. We found that the difference between the generated final result (**Figure 2i**) and the real BF-TEM image (**Figure 2g**) is very small, and the segmentation network trained with virtual data also achieves the requirement of recognizing all nanoparticles with excellent performance. Our findings indicate that pre-trained models from different domains integrate well into the tGAN pipeline, effectively utilizing transfer learning.

This approach enables the training of highly accurate semantic segmentation models without manual labeling, significantly advancing the field of automated EM image analysis.

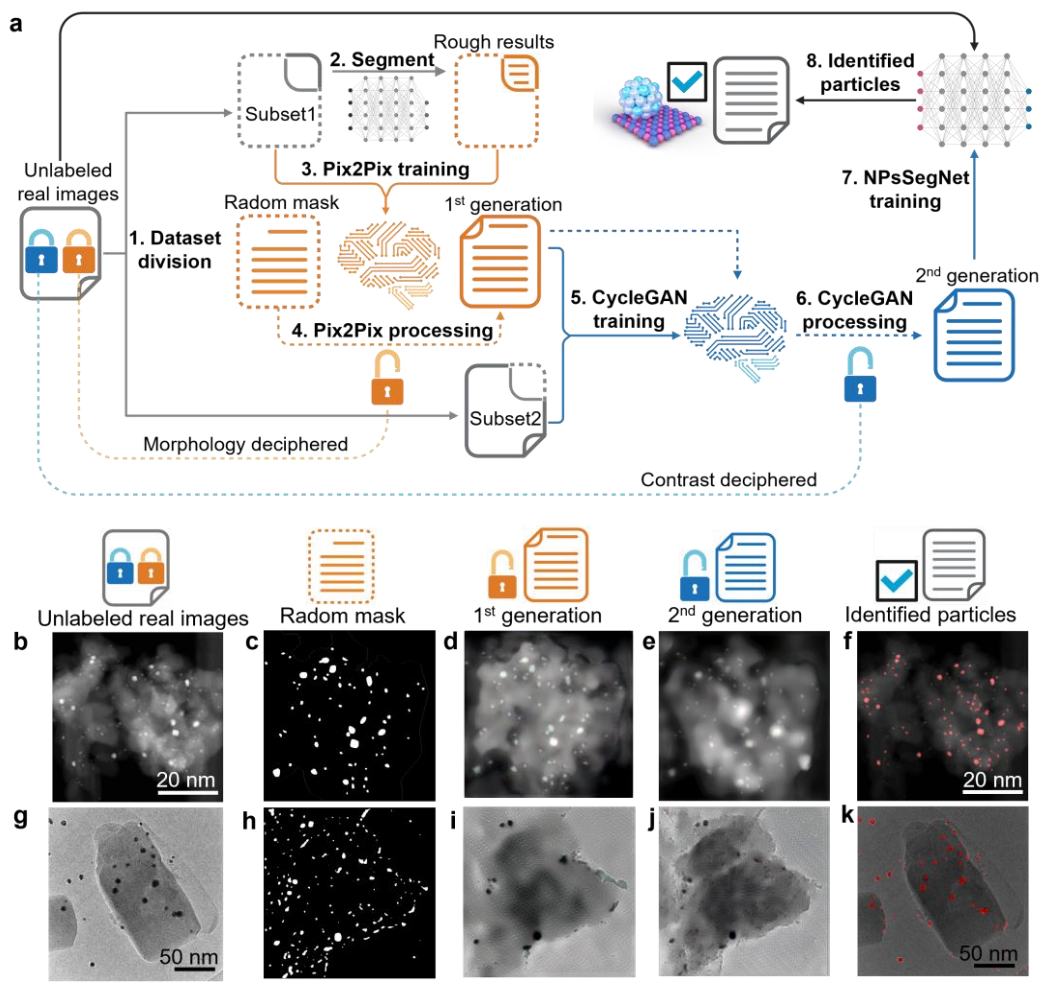

**Figure 2 | Workflow of the proposed tandem generative adversarial network (tGAN) approach.** (a) Schematic pipeline of data generation and segmentation network training. (b-f) Representative HAADF-STEM images stream throughout the workflow, including (b) original HAADF image, (c) mask, (d) 1$^{st}$ generation, (e) 2$^{nd}$ generation, and (f) segmentation result . (g-k) Representative BF-TEM images stream throughout the workflow, including (g) original TEM image, (h) mask, (i) 1$^{st}$ generation, (j) 2$^{nd}$ generation, and (k) segmentation result .

**Ablation Studies for Morphology and Contrast**

After determining the generalizability of tGAN, the rationality of the design of the tGAN generation pipeline is another issue of concern to us, so we further evaluated whether it has the tandem nature of successively acquiring the morphology and contrast information. As shown in **Figure 3a**, as the size of the pix2pix training dataset increases (from 5 to 72 images), the FID scores of the 1st generation compared to the real images drop and the IS scores go up, suggesting that its similarity and diversity have improved, which is consistent with common sense. But most importantly, if we keep using the smaller dataset (5 images, which is typically the number of high-quality images acquired for a new sample in a single HAADF-STEM characterization slot) to train pix2pix, but then attach CycleGAN as a relay to produce the 2nd generation, the similarity and diversity of the data are even higher than the result of just boosting the amount of data for pix2pix. So, through this phenomenon we found that the existence of CycleGAN can well compensate for the low data utilization in the presence of only pix2pix, *i.e.*, it is precisely this tandem architecture that allows tGAN to fully decipher the information convoluted in the data. The same phenomenon is verified for the SSIM and PSNR metrics (**Figure 3b**), again confirming that this tandem design maximizes the quality of the generated virtual dataset. To quantitatively assess the morphological and contrast features extracted by two GANs. We adopted the kernel MMD metric to quantify the improvements in contrast features through CycleGAN. Because of MMD's excellent ability to discriminate noisy pictures in the RestNet feature space[50], and precisely the contrast information of EM images is generally responded to by the degree of noise at the edges of the particles, it is reasonable to believe that MMD can better reflect the contrast features compared to other metrics. As shown in **Figure S1**, when the pix2pix model is used alone, the data volume has very little effect on the MMD (in contrast to metrics of FID and SSIM, which represent more in morphology). Once the CycleGAN is added in series, the MMD value decreases significantly, indicating that the generated image is much closer to the real image and confirming the sensitivity of CycleGAN to contrast features. Furthermore, by visualizing the representative images shown in **Figure 3c**, it can also be noticed that only the results with the addition of CycleGAN are the most realistic. Therefore, the tandem nature of tGAN ensures efficient data utilization, which in turn helps the realization of label-free.

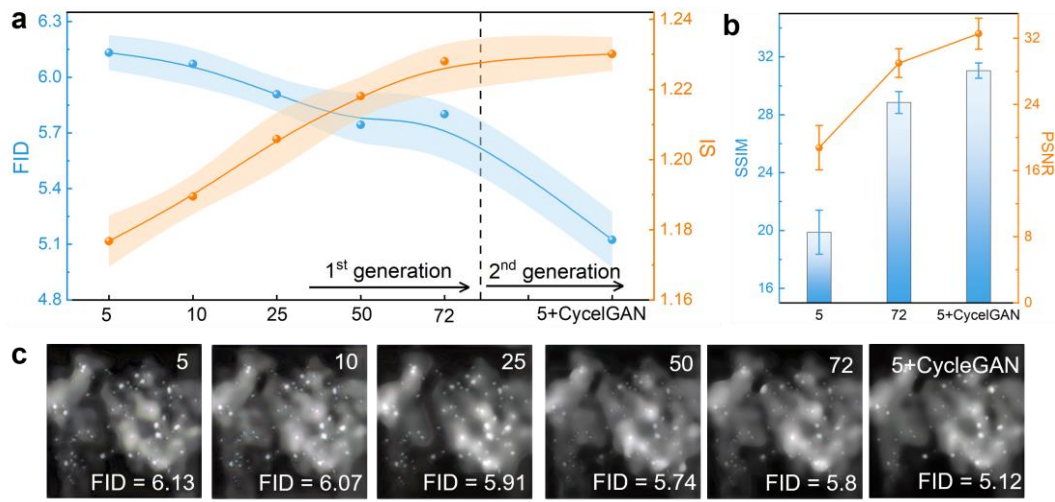

**Figure 3 | Evaluation of the quality of generated virtual dataset.** (a) Fréchet inception distance (blue) and inception score (orange) of generated datasets, the labels of x-axis are the number of images used for model training. (b) Structure similarity index measure and peak signal-to-noise ratio of generated datasets. (c) Representative generated images from different datasets.

**Validation of the tGAN Model**

We chose NPsSegNet model as the segmentation model for validation in this work, as NPsSegNet has showed superior performance with higher mean MIoU (**Figure S2**). Particularly, it is more effective in segmenting smaller sized particles with poor contrast (**Figure S3**). This can be attributed to the more intricate architecture of UNet++, which is better suited for capturing fine-grained details. Furthermore, the statistical analysis indicates that the improvements in performance metrics (PA and MIoU) of NPsSegNet over UNet and DeepLabV3+ are statistically significant (**Table S2**), meaning that the observed differences are unlikely to have occurred by chance and can be attributed to the superior design of NPsSegNet.

To ascertain whether the generated virtual data can be utilized for NPsSegNet, we quantified the training performance of the label-free method based on tGAN and the manually-labeling method. As shown in **Figure 4a**, the convergence process of NPsSegNet trained with virtual data is faster compared to manual-labeling data. More importantly, the pixel accuracy (PA) of the label-free method surpasses that of the manual-labeling method, as illustrated in **Figure 4b**. This indicates higher data utilization efficiency with the label-free method. Moreover, compared to the manual-labeling method, which inevitably introduces biases during training, the automatically generated labels are more accurate, resulting in a superior segmentation inference. To further validate this conclusion, we employed a more accurate evaluation metric, Mean Intersection over Union (MIoU), as shown in **Figure 4c**. This also indicates the higher accuracy of the label-free method. All the evidence above suggests that the label-free method is more accurate than the manual-labeling method. The detailed differences between the results of these two methods can be observed in **Figures 4d-f**. **Figure 4d** shows a representative HAADF real image, while **Figures 4e and 4f** depict the results of the label-free method and manual-labeling method, respectively. It is evident that the label-free method can identify more nanoparticles compared to the manual-labeling method, demonstrating the superiority of its model. However, when confronted with particles of extremely poor contrast (*e.g.*, out-of-focus), the label-free method is still less accurate, probably because the dataset used for training is still not comprehensive enough to cover all possible contrast scenarios, leaving room for improvement here in the future. Additionally, tGAN-based NPsSegNet has been deployed for on-the-fly analysis (Supplementary Video 1), providing a one-stop solution for the stringent requirements of model generalizability in *in-situ* experiments due to the variable experimental parameters and allowing real-time feedback.

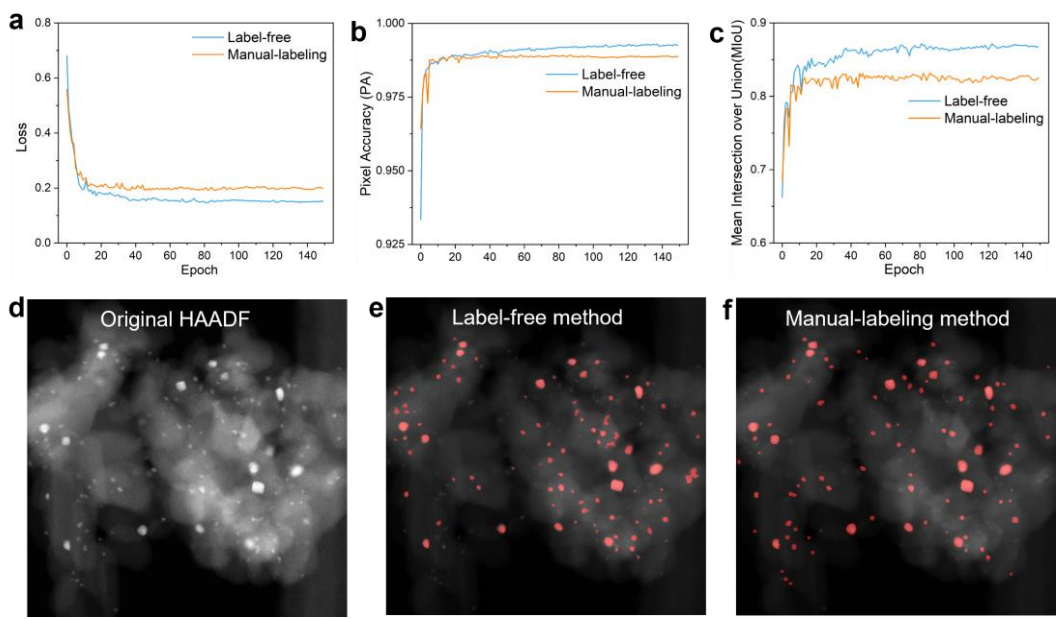

**Figure 4 | Evaluation of segmentation network trained by label-free method and manual-labeling method.** (a) Loss curve, (b) pixel accuracy, (c) mean intersection over union of label-free method and manual-labeling method. (d) A representative HAADF-STEM image of PtSn@Al$_2$O$_3$. (e)Segmentation result of (d) using label-free method. (f) Segmentation result of (d) using manual label method.

**Conclusion**

This study introduces a novel tandem generative adversarial network (tGAN) that achieves simulation-free and label-free generation of realistic EM images. Our results show that tGAN pipeline, which deciphers morphology and contrast features sequentially, significantly enhances the quality and generalizability of generated datasets. Evaluation metrics of FID, IS, SSIM, and PSNR confirm that tGAN outperforms conventional GANs. The label-free method using tGAN demonstrated superior performance over manually-labeling methods in training the nanoparticles segmentation network (NPsSegNet), with faster convergence and higher segmentation accuracy (via PA and MIoU). This approach mitigates biases inherent in manual labeling, offering more robust and reliable segmentation. Future work could explore the application of tGAN in other imaging modalities and further enhance model generalizability across diverse datasets. Integrating tGAN-based NPsSegNet for real-time analysis in in-situ experiments advances model generalizability under varied conditions, providing efficient and accurate data analysis. In a nutshell, this framework addresses limitations of conventional labeling methods and sets a new standard for virtual dataset generation in the EM flied, with potential applications of imaging analysis tools in other scientific fields.

## Acknowledgments

Q. He acknowledges the support from National Research Foundation (NRF) Singapore, under its NRF Fellowship (NRF-NRFF11-2019-0002).

## Author contributions

Q. He and W. Yuan co-conceived the research idea. W. Yuan designed and conducted the experiments and coding. B. Yao and S. Tan performed the (S)TEM characterizations and provided the microscopy data. W. Yuan and Q. He drafted the manuscript, and F. You contributed to extensive reviews and revisions. All the co-authors contributed to the discussion and commented on the manuscript.

## Competing interests

For the software in this work, the EMcopilot® has been registered as a trademark by the Intellectual Property Office of Singapore (IPOS, NO.40202319264T). The authors declare potential economic interests associated with the usage of the EMcopilot® trademark in respect of the service in Class 42 (Software as a service, SaaS).